\documentclass[%
rsi,%
superscriptaddress,
 amsmath,amssymb,
 a4paper,11pt%
]{article}

\usepackage{jcappub} 

\usepackage[T1]{fontenc} 

\usepackage{graphicx}
\usepackage{dcolumn}
\usepackage{bm}
\usepackage{color}
\usepackage{appendix}

\title{Cosmic no-hair theorems for viscous contracting Universes}
\author{Chandrima Ganguly}
\affiliation{Wolfson College, Barton Road, Cambridge CB$3$ $9$BB}
\affiliation{DAMTP, Centre for Mathematical Sciences, University of Cambridge Wilberforce Road, Cambridge CB$3$ $0$WA, United Kingdom}
\emailAdd{c.ganguly@damtp.cam.ac.uk}

\abstract{A cosmic no-hair theorem for all initially contracting, spatially homogeneous, orthogonal Bianchi Cosmologies is derived - which shows that all such Universes asymptote to a spatially flat, isotropic Universe with the inclusion of a shear viscous stress. This establishes a new mechanism of isotropisation in a contracting Universe, which does not take recourse to an ekpyrosis-like mechanism using an effective ultra-stiff equation of state fluid, that is, one in which the pressure is much greater than the energy density.}

\begin{document}

\maketitle
\flushbottom

\section{Introduction}

The inflationary paradigm \cite{inflation} provides a simple framework to generate primordial nearly scale-invariant perturbations with a slightly red tilt, as well as small non-gaussianties which are in good agreement with observations \cite{Planck1,Planck3,Planck4}. Despite its successes, the inflationary scenario has certain issues \cite{problems-ijjas-inflation,trans-Planckian} within its models and there is much discussion within the early universe cosmology community around attempts to fix these issues within the paradigm. The other approach is to try to construct an alternative paradigm \cite{Gasperini2002,brand} to see if this can also explain the observations of the late-time Universe that we have today. The bouncing cosmology paradigm is one such attempt \cite{review,review2}. 

\paragraph*{}
We will be interested in model realisations within this paradigm that are non-singular. This means that these models do not suffer from the initial Big Bang singularity that is ubiquitous in the case of cosmologies that have expanded for its entire history \cite{vilenkin-borde-guth}. The trade-off in the case of a bouncing model is that there has to be some new physics at the expansion minimum (which coincides with the `bounce') to allow the Universe to re-expand \cite{Lehners:2007ac,Lehners:2008vx,Lehners2008}. This new physics is in the form of a null energy condition violation in the case of spatially flat models. The inclusion of positive curvature does satisfy one of the conditions on the vanishing of the Hubble rate without the need of violating the null-energy condition \cite{BarrDab}. However a second condition on the successful fulfillment of the bounce is to have the first derivative of the Hubble rate be negative and some new physics would be needed to have the Universe re-expand from the minimum. In a previous work, we have parametrised this `new physics' by hypothesising a non-linear equation of state \cite{me4}. This non-linear equation of state was used in \cite{bruni1,bruni2} to model a dynamical dark energy scenario. This model has the advantage that there exists a high energy cosmological constant which represents the maximum value the energy density can take without causing the model to exhibit phantom behaviour. At this energy density, the null energy condition is not violated at the level of phenomenology. This means that in the presence of positive spatial curvature it is still possible to have a bounce with no violation of the null-energy condition, and only a violation of the strong energy condition.

\paragraph*{}
An additional problem that plagues contracting universe scenarios, such as one that must be incorporated in a bouncing cosmology, is the problem of growing anisotropies and inhomogeneities. The issue with growing anisotropies is not only that this is in direct contradiction to the very tight constraints on anisotropy \cite{saadeh} and the time of isotropisation in the late Universe - but also that a successful bounce must necessarily have very low anisotropy going into it from the contracting phase \cite{ekpyrosis_mixmaster,ekpyrosis_numerics}. For example, it has been shown \cite{review} that at the level of perturbations if the mechanism for example is a scalar field $\chi$ that is causing the bounce to occur, then the perturbative anisotropy energy density $\sigma ^2$ must obey the ratio $\frac{\dot{\chi}^2}{\sigma ^2} \gg 1$ for the bounce to occur successfully and lead to re-expansion. This can be understood non-perturbatively as well - a contracting closed anisotropic Universe suffers an instability known as a BKL instability by which the Universe undergoes infinite chaotic oscillations on a finite time interval until it finally collapses to a singularity \cite{mixmaster_numerics_1,mixmaster_numerics_2,bkl,belinski_henneaux_2017,Belinskii1972}. The anisotropies thus need to be suppressed to avoid this instability from being manifested in a contracting Universe.

\paragraph*{}
A way of dealing with the unbounded growth of anisotropies has been to incorporate an ekpyrotic phase \cite{Lehners:2007ac,Lehners:2008vx,Lehners2008,review,review2}. This is described by an `ultra-stiff' equation of state i.e.\ an equation of state that is given by $P\gg \rho$ where $P$ and $\rho$ denote the pressure and energy densities of the fluid. In the ekpyrotic scenario, this is incorporated with the aid of a slow contraction mediated by a fast-rolling scalar field rolling down a negative exponential potential. While this model has been proved to be very successful in dealing with the BKL instability \cite{mixmaster_numerics_1,mixmaster_numerics_2,ekpyrosis_mixmaster}, it still has some issues. For example, it seems to be a fine-tuned scenario when one incorporates the effects of anisotropic pressures even if these follow the condition of ultra-stiffness on average \cite{me1}. It also has a problem of perturbations being blue-tilted without the incorporation of an additional scalar field \cite{review,review2}. There is also the probability of super-luminal propagation of the kinetic energy modes of the scalar field that is an issue in any scenario that has $p>\rho$. All of these reasons have prompted us to search for an alternative mechanism for isotropising a contracting Universe. It has been shown in \cite{me4,misner_aniso} that the incorporation of a negative anisotropic stress in the form of shear viscosity such that the anisotropic stress is proportional to the shear anisotropy with a negative proportionality constant leads to the isotropisation of a Bianchi IX Universe.

\paragraph*{}
In this work we shall study the global stability of the isotropic Friedmann-Lemaitre point in the absence of the ultra-stiff fluid equation of state and in the presence of this shear viscous stress. We shall show that a contracting Universe sourced by a non-linear equation of state and with the incorporation of shear viscosity demonstrates an attractor behaviour on approach to this isotropic Friedmann-Lemaitre point. In doing so, we are drawing on previous work on no-hair theorems in contracting universes \cite{Lidsey2005} to derive a new no-hair theorem which shows that isotropisation with the aid of shear viscous anisotropic stress is possible in the most general spatially homogeneous Universe. We do not model the bounce in this work. In order to do this, we would need a specific geometry as it would depend on whether the model in question was flat or had non-zero spatial curvature. Our aim remains to derive a general result for the process of isotropisation in the presence of shear viscosity in a contracting Universe. 

\paragraph*{}
In the next section, we shall provide a brief background on Bianchi cosmologies which shall serve as the framework for the remainder of the paper. In the subsequent sections, we shall use the expansion normalised variables of the orthonormal formalism as done in \cite{WEllis,Lidsey2005} to prove a general cosmic no-hair theorem for the isotropisation of the Bianchi cosmologies in the presence of shear viscous anisotropic stresses for Bianchi Types I-VIII, Type IX and separately Bianchi Class B.

\paragraph*{}
 For the entirety of this work we have used natural units where $c = \hbar = G =1$.

\section{A background on Bianchi cosmologies}

We are, in this work, interested in studying the stability of the isotropic Friedmann Lemaitre point. Thus we employ the class of the most general spatially homogeneous, anisotropic cosmologies which are given by the Bianchi cosmologies. The summary in this section will largely follow the formalism as laid out in \cite{WEllis}.
These admit a local group $G_3$ group of isometries that act simply transitively on a spacelike hypersurface. Then the line element can be written in the form $ds^2 = -dt^2 +h_{ab}\omega^a \omega ^b$ where $\omega ^a$ are the $1$-forms. These $1$-forms follow the Maurer-Cartan equations where $d\omega ^c =\frac{1}{2}C^c_{ab} \omega ^a \wedge \omega ^b$ where $C^c_{ab}$ are the structure constants of the Lie algebra. They are anti-symmetric so $C^c_{(ab)}=0$. This means that the independent components of the structure constants can be expressed as a symmetric $3\times 3$ matrix $n_{ab}$ and the components of a $3\times 1$ vector $A_b =C^a_{ab}$. So the structure constants can be written as,
\begin{equation}
    C^c_{ab} = n^{cd}\epsilon_{dab} + \delta^c_{[a}A_{b]}
\end{equation}
where $\epsilon_{abc}$ is the antisymmetric tensor. Using the Jacobi identity, 
\begin{equation}
    C^e_{d[a}C^d_{bc]}=0
\end{equation}
we find that,
\begin{equation}
    n_{ab}A^a =0
\end{equation}
We use the freedom of choice of orthonormal frame to rotate the frame vectors and diagonalise the tensor $n_{ab}$ as 
\begin{equation}
    n_{ab} = \mathrm{diag}\left\{n_1,n_2,n_3\right\}
\end{equation}
If $A^a \neq 0$ then we see that the above is an eigenvalue equation with the eigenvector $A^a$ always being able to be expressed as $A^a = (A,0,0)$. Therefore the eigenvalue equation simplifies to,
\begin{equation}\label{eq:paramAB}
    n_1A =0
\end{equation}
\textit{For Bianchi Class A, $A=0$ and for Bianchi Class B $A\neq 0$}. We also assume that the direction of fluid flow is orthonormal to the group orbits. Hence we use the orthonormal frame approach and therefore we can write out the Einstein Field Equations in all generality for the Bianchi Classes A and B. The fluid velocity which is the unit timelike vector is given by $u_a$. This defines a projection tensor for the metric on the surface $g_{ab}$ given by,
\begin{equation}
    h_{ab} = g_{ab} +u_a u_b
\end{equation}
which at each point projects onto the $3$-space orthogonal to the vector $u_a$ \cite{WEllis}.
The generalised expansion scalar is given by $\Theta$
\begin{equation}
    \Theta = u^a_{;a}
\end{equation}
 $\Theta= 3H$ where $H$ is the Hubble rate. 
The evolution equation for the fluid velocity is given by,
\begin{equation}
    \dot{u}_a = u_{a;b}u^b
\end{equation}
In this notation, the shear tensor is given by,
\begin{equation}
    \sigma_{ab} = u_{(a;b)}-\frac{1}{3}\Theta h_{ab}-\dot{u}_a u_b
\end{equation}

The evolution equation for the shear is given by,
\begin{equation}\label{eq:evshear}
    \dot{\sigma}_{ab } = -3 H\sigma _{ab} +2 \epsilon ^{m n}_{(a}\sigma_{b)m}\Omega_{n}- \phantom{p}^3 S_{ab} +\pi_{ab}
\end{equation}
The shear $\sigma _{\alpha \beta}$ can be diagonalised in the frame in which $n_{\alpha \beta}$ is diagonal and by the requirement of tracelessness has only $2$ independent components. We shall work with some linear combinations of these components as,
\begin{equation}
    \sigma _+ =\frac{1}{2}\left(\sigma _{22}+\sigma_{33}\right),\;\;\; \sigma _- =\frac{1}{2\sqrt{3}}\left(\sigma _{22}-\sigma_{33}\right)
\end{equation}
The other unknown quantity in \eqref{eq:evshear} is the local angular velocity of the spatial frame specified by a set of $1$-forms $\{\textbf{e}_{\alpha}\}$ with respect to a Fermi-propagated spatial frame. We shall call this the Fermi rotation which is given by,
\begin{equation}
    \Omega ^{a} = \frac{1}{2}\epsilon^{a m n}e^i_{m}e_{n i;j}u^j
\end{equation}

As mentioned before, we can choose a frame where $n_{\alpha \beta}$ are diagonalised and the  $A_{a}$ can be expressed as $(A,0,0)$ via the Jacobi equation. We can show that the shear tensor can also be diagonalised in this frame as long as $\pi_{\alpha \beta} =0$ or $\pi_{\alpha \beta} \propto \sigma_{\alpha \beta}$. Details of this proof can be found in the appendix.
 Thus the Einstein field equations in the orthonormal system formalism become,
 \begin{align}
 \dot{\sigma}_{\pm}&=-3H\sigma_{\pm} - ^{(3)}S_{\pm}\\
 \dot{n}_1 &=\left(-H-4\sigma _+\right)n_1\\
 \dot{n}_2 &= \left(-H+2\sigma_+ + 2\sqrt{3}\sigma _-\right)n_2\\
 \dot{n}_3 &= \left(-H+2\sigma_+ -2\sqrt{3}\sigma _-\right)n_3
 \end{align}
 along with the continuity equation,
 \begin{equation}
     \dot{\rho}=-3H\left(\rho +p\right) -\left(\pi_+\sigma_+ + \pi_-\sigma_-
     \right)
 \end{equation}
 where $\pi_{\pm}$ is given by,
 \begin{equation}
     \pi_{\pm} = -\kappa H \sigma_{\pm}
 \end{equation}
 In subsequent sections, we shall express these equations as a phase plane system through the process of expansion normalisation. In the case of Bianchi Types I-VIII we can normalise the gravitational field variables with respect to the Hubble expansion rate and in the case of Type IX which has a positive spatial curvature, this normalisation occurs with respect to a modified expansion variable. 
 \paragraph*{}
 Before we begin our analysis of the stability of the isotropic Friedmann-Lemaitre point, we shall review the inclusion of dissipative effects in a cosmology.

 \section{Shear and Bulk viscosity}
 In standard cosmological analysis, dissipative effects are ignored. However their inclusion can give rise to many important effects - such as singularity resolution or novel isotropisation mechanisms.
 \paragraph*{}
Dissipative effects can be modelled by including shear and bulk viscosity of the component fluids in the cosmology. This has been studied in several works \cite{gron-viscous}. The effect of bulk viscosity has been to increase the expansion rate i.e.\ modify the volume expansion. The manifestation of this is that the pressure of the component fluid gets modified as $\tilde{p} = p -3\xi H$ where $H$ is the Hubble expansion and $\xi$ is the coefficient of bulk viscosity. For bulk viscosity arising from radiation fluids, this effect vanishes when the equation of state of the radiation fluid is $p=(1/3)\rho$ i.e.\ the relativistic limit. However, viscosity effects can become important arbitrarily early in the history of the Universe if these effects were generated by particle collisions from gravitons, or processes like the evaporation of mini black holes. Bulk viscosity has also been studied as a mechanism of singularity resolution in cosmologies - for example for certain initial conditions, flat FLRW models with non-linear viscous terms are shown to be able to admit non-singular solutions. The non-linear equation of state that has been studied in \cite{bruni1,bruni2, me4} can be modelled as a bulk viscosity that is a function of the expansion rate $H$ - in the cases where the curvature is zero. Thus it is not surprising that non-singular solutions are found in the context of these non-linear fluids \cite{bruni1}.
 \paragraph*{}Shear viscosity, on the other hand, usually manifests as an anisotropic pressure term and modifies the tensor modes, or the anisotropies of the cosmology. This has been studied with respect to neutrinos in \cite{misner_aniso} and also in the case of the early Universe for warm inflationary models \cite{gil}. It has been used to resolve the anisotropy problem in bouncing cosmologies in the contracting phase in \cite{me4}. For the case of small anisotropy, the form of shear viscosity is derived in \cite{weinberg_viscosity}. The anisotropic pressure term derived from shear viscosity there is given by $-\tilde{\kappa} \sigma_{ab}$ where $\tilde{\kappa}$ is the coefficient of viscosity. In \cite{belinski_henneaux_2017}, the coefficient of viscosity is dependent on density. When $\tilde{\kappa} \sim \rho ^n$, for values of $n < 1/2$, the Friedmann singularity for a contracting Universe is unstable and acausal. For $n>1/2$, the Friedmann singularity forms but is unstable. The only stable isotropisation on contraction occurs at $n=1/2$. Hence to achieve effective isotropisation, we choose the shear viscosity to take the form $-\kappa \rho ^{1/2} \sigma _{ab}$. For Universes that are flat and isotropic, this is the same as $-\kappa H \sigma_{ab}$. This is not the case in all of the general Bianchi types that we will be discussing in this work. However this choice makes the process of expansion normalisation and hence the phase plane analysis tractable. This form of the shear viscous term also makes the modified coefficient of viscosity $\kappa$ dimensionless. This simplification would not make a difference to the isotropising effect of shear viscosity. We can demonstrate this by considering the most anisotropic spatially homoegeneous geometry in a situation of anisotropy domination - the anisotropy dominated Kasner Universe. In this scenario, $H\sim \sigma ^2 \sim \sigma_{ab}\sigma^{ab}$. The evolution equation for the shear in the presence of viscous stress as modelled by $-\kappa H \sigma_{ab}$ is given by,
 \begin{equation}
     \dot{\sigma}_{ab} +3 H\sigma_{ab} = -\kappa H \sigma_{ab}
 \end{equation}
Using the limit of anisotropy domination, we find the solution to the anisotropy energy density $\rho_{\sigma}$ which is proportional to $\sigma ^2 \sim \sigma_{ab}\sigma^{ab}$ given by,
\begin{equation}
   \rho_{\sigma} \sim (6t +2 kt -c_1)^{-2}
\end{equation}
 If the singularity which would turn into the bounce in the presence of positive curvature or new physics occurred at $t\rightarrow -\infty$ then $\rho_{\sigma}\rightarrow 0$ at this point. So the shear viscous term is successful in isotropising a Universe with maximum possible anisotropy i.e.\ the anisotropy-domanted Kasner Universe.
 
 \section{Cosmic no-hair theorems}

 In this section we follow the procedure laid out in \cite{Lidsey2005} to prove a global stability theorem for the most general, spatially homogeneous cosmologies. We consider a fluid with an equation of state given by,
 \begin{equation}
     p = \left(\gamma(\rho)-1\right)\rho
 \end{equation}
 Our results hold for a general $\gamma(\rho)$ - for the ideal equation of state, $\gamma$ is a constant. In \cite{Lidsey2005} they had explored the possibility of using an ultra-stiff equation of state i.e.\ $\gamma >2$ to prove a general cosmic no-hair theorem about the isotropic FL point being an attractor for initially contracting Universes of Bianchi Classes A and B. We use the same methods to prove a similar theorem - but this time in the absence of an ultra-stiff equation of state. Our ingredient is a shear viscous term - further cementing the idea that such a viscous effect can be used to isotropise Universes without necessarily having to take recourse to an `ekpyrosis'-like ultra-stiff fluid  \cite{me4}. We have simply required that the Strong Energy Condition be obeyed, i.e.\ $\gamma >2/3$. 
We define a new time variable $\tau$ which is defined as
\begin{equation}
    \frac{dt}{d\tau}=\frac{1}{H}
\end{equation}
The time derivatives are taken with respect to this variable $\tau$ and is denoted by $'$. The Big Crunch singularity would take place at $\tau \rightarrow -\infty$. For a bounce model, this singularity would be replaced by a bounce and subsequent re-expansion. This would require new physics at the expansion minimum. For the case of a non-linear model \cite{bruni1,bruni2}, the presence of positive curvature and the non-linear equation of state fluid satisfies the conditions $H=0$ and the derivative of the Hubble rate with respect to cosmic time, $\dot{H}>0$ required for the bounce to occur. For the purposes of this work, we have not modelled the bounce but are instead interested in isotropisation in the contracting phase.
\paragraph*{}
 As we are interested in the dynamics of the contracting phase alone, we can write out the Einstein's equations in expansion normalised form.

\begin{align}
     \dot{\Sigma}_{\pm}' &=-(2-q)\Sigma_{\pm} -S_{\pm}\\
     \dot{N}_1'&=\left(q-4\Sigma_+\right)N_1\\
     \dot{N}_2'&=\left(q+2\Sigma_++2\sqrt{3}\Sigma_-\right)N_2\\
     \dot{N}_3'&=\left(q+2\Sigma_+-2\sqrt{3}\Sigma_-\right)N_2\\
 \end{align}
 
where
\begin{equation}\label{eq:defOfq}
    q = 2\Sigma ^2 +\frac{1}{2}\left(3\gamma -2\right)\Omega
\end{equation}
 and the expansion normalised variables are given by,
 \begin{equation}\label{eq:normalisationI-VIII}
    \left(\Sigma_+,\Sigma_-,N_1,N_2,N_3,\Omega\right)=\left(\frac{\sigma_+}{H},\frac{\sigma_-}{H},\frac{n_1}{H},\frac{n_2}{H},\frac{n_3}{H},\frac{\rho}{3H^2}\right)
\end{equation}

The Friedmann constraint is given by,
\begin{equation}\label{eq:Friedmann_constraint}
    \Sigma ^2 +\Omega + K =1
\end{equation}

Here $K$ and $\Sigma ^2$ are respectively given by,
\begin{equation}\label{eq:defK}
    K =\frac{1}{12}\left(N_1^2+N_2^2+N_3^3 -2N_1N_2-2N_2N_3-2N_3N_1\right)
\end{equation}
\begin{equation}\label{eq:defsigma}
    \Sigma^2 = \Sigma _+^2+\Sigma _-^2
\end{equation}
which for Types I-VIII, is a positive definite quantity.

 Let us look at the expansion normalised evolution equation for the density $\Omega$,
 \begin{equation}
     \Omega ^{\prime} = \left[ -(3\gamma -2)K +3(2-\gamma)\Sigma ^2\right]\Omega
 \end{equation}
In the case of an ultra-stiff fluid $\gamma >2$, we can see that under the condition that $K>0$ , $\Omega  ^{\prime} \leq 0$ for any initially contracting Bianchi $I-VIII$ Universe, where equality occurs iff $K=\Sigma ^2 =0$ for any non-vacuum orbit $\Omega\neq 0$. As $\Omega$ is a monotonic decreasing function of $\tau$ and it is bounded by the Friedmann constraint, we can infer that $\lim_{\tau \to -\infty}\Omega^{\prime}=0$. This further implies that,
\begin{equation}
    \lim_{\tau \to -\infty}K=0,\; \lim_{\tau \to -\infty}\Sigma =0,\;\lim_{\tau \to -\infty}\Omega = 1
\end{equation}
Thus this means that for an initially contracting Bianchi $I-VIII$ Universe, isotropisation must have occurred as the bounce point is approached. If we now have a situation where the anisotropic stress is non zero and is of the form,
\begin{equation}
    \pi _{ab} = -\kappa H \sigma_{ab}
\end{equation}
we get the evolution equation for $\Omega$ as,
\begin{equation}\label{eq:omegaEvEq}
     \Omega ^{\prime} = -(3\gamma -2)K\Omega +\Sigma ^2 \left[3(2-\gamma)\Omega -\kappa\right]
\end{equation}

For our quadratic EOS, we have\cite{me4} 
\begin{equation}
    \gamma = \frac{P}{\rho} +1 = (\alpha+1) -\beta \frac{\rho}{\rho_C}
\end{equation}
 We use the expansion normalised Friedmann constraint \eqref{eq:Friedmann_constraint} to write the evolution equation for $\Omega$ \eqref{eq:omegaEvEq} as follows,
\begin{equation}\label{eq:EvModOmega}
     \Omega ^{\prime} = -(3\gamma -2)K \Omega +\left[\left\{3(2-\gamma)-\kappa\right\}\Omega -\kappa \Sigma ^2 -\kappa K\right]\Sigma^2
\end{equation}
If the strong energy condition i.e.\ $\rho + 3p\geq0$ is obeyed, then $(3\gamma -2)$ in the first term of \eqref{eq:EvModOmega} is positive definite. If $\kappa>0$, then using the fact that $\Sigma ^2 >0$ and $K>0$, we have the condition on  $\Omega  ^{\prime} \leq 0$ as
\begin{equation}\label{eq:condition}
    \kappa > 3\left(2-\gamma\right)
\end{equation}
The case in which $\kappa =0$, for $\Omega ^{\prime} \leq 0$, we need the equation of state to be ultra-still effectively i.e.\ $\gamma(\rho)>2$.
If $\Omega ^{\prime} \leq 0$ for initially contracting Bianchi Types $I$-$VIII$ with equality when $K=\Sigma ^2 =0$, this means that $\Omega$ is a monotonically decreasing function of $\tau$ and as $\Omega$ is bounded by the Friedmann constraint \eqref{eq:Friedmann_constraint}.
we can conclude that $\lim_{\tau \rightarrow -\infty}\Omega ^{\prime} =0$. This implies that $\lim_{\tau \rightarrow -\infty}K=\lim_{\tau \rightarrow -\infty}\Sigma ^2 =0$ and by \eqref{eq:Friedmann_constraint}, $\lim_{\tau \rightarrow -\infty}\Omega =1$. Using $\Sigma^2 = \Sigma _+^2+\Sigma _-^2$, we can conclude that $\lim_{\tau \rightarrow -\infty}\Sigma _+ =\lim_{\tau \rightarrow -\infty}\Sigma _-=0$. This also implies that $\lim_{\tau \rightarrow -\infty}q=(3\gamma -2)/2>0$ for fluids obeying the strong energy condition. The evolution equations for the expansion normalised curvature variables are as follows,
\begin{align}
  N_1 '&=\left(q-4\Sigma_+\right)N_1\\
  N_2'&= \left(q+2\Sigma_+ +2\sqrt{3}\Sigma_-\right)N_2\\
  N_3'&= \left(q+2\Sigma_+ -2\sqrt{3}\Sigma_-\right)N_3\\
\end{align}
Using the value of $q$ at $\tau \rightarrow -\infty$ we see that there exists a parameter $\epsilon >0$ such that $N'_i/N_i >\epsilon$ for a sufficiently negative $\tau$ so that $\lim _{\tau \rightarrow -\infty}N_i =0$. So following the same reasoning as in \cite{Lidsey2005}, we can conclude that for the values of $\kappa$ obeying \eqref{eq:condition}, the isotropic solution is a global attractor for a contracting Universe.

\section{Bianchi IX Cosmic No Hair Theorem}

The analysis for Bianchi IX has to be done slightly differently to the Bianchi Types I-VIII above. This is because the curvature variables $n_1$, $n_2$, $n_3$ are all positive in this Bianchi type and hence the spatial curvature is also positive. The Bianchi Type IX is an anisotropic generalisation of the closed Friedmann Universe. This also means that the expansion is no longer monotonic and the Hubble rate $H$ can be zero. Thus expansion normalisation of the state vector $\textbf{x} = \left \{ H,\sigma_+,\sigma_-,n_1,n_2,n_3\right\}$ must be done with respect to a modified expansion variable,
\begin{equation}\label{eq:defOfD}
    D = \sqrt{H^2 + \frac{1}{4}\left(n_1 n_2 n_3\right)^{2/3}} 
\end{equation}
The normalisation then becomes,
\begin{align}\label{eq:normalisationIX}
    \left(\bar{H},\bar{\Sigma}_+,\bar{\Sigma}_-,\bar{N}_1,\bar{N}_2,\bar{N}_3,\bar{\Omega}\right)\equiv\\ \nonumber
    \left(\frac{H}{D},\frac{\sigma_+}{D},\frac{\sigma_-}{D},\frac{n_1}{D},\frac{n_2}{D},\frac{n_3}{D},\frac{\rho}{3D^2}\right)
\end{align}
From \eqref{eq:defOfD} and \eqref{eq:normalisationIX}, we see that $\bar{H}$ is constrained by,
\begin{equation}\label{eq:defHbar}
    \bar{H}^2 +\frac{1}{4}\left(\bar{N}_1 \bar{N}_2 \bar{N}_3\right)^{2/3} =1
\end{equation}
In correspondence to the Types I-VIII, we must define a new time variable $\bar{\tau}$,
\begin{equation}
    \frac{dt}{d\bar{\tau}}=\frac{1}{D}
\end{equation}
The evolution equations then become
\begin{align}
    D^{\star}&=-(1+\bar{q})\bar{H}D\\
    \bar{\Sigma}_{\pm}^{\star} &= -(2-\bar{q})\bar{H}\bar{\Sigma}_{\pm}-\bar{S}_{\pm}\\
    \bar{N}_1^{\star}&=\left(\bar{H}\bar{q}-4\bar{\Sigma}_+\right)\bar{N}_1\\
    \bar{N}_2^{\star}&=\left(\bar{H}\bar{q}+2\bar{\Sigma}_++2\sqrt{3}\bar{\Sigma}_-\right)\bar{N}_2\\
    \bar{N}_3^{\star}&=\left(\bar{H}\bar{q}+2\bar{\Sigma}_+-2\sqrt{3}\bar{\Sigma}_-\right)\bar{N}_3
\end{align}

Here the quantity $\bar{q}$ is defined by
\begin{equation}\label{eq:defOfqBIX}
    \bar{q}= \frac{1}{2}(3\gamma -2)(1-\bar{V})+\frac{3}{2}(2-\gamma)\bar{\Sigma}^2
\end{equation}
The normalised Friedmann constraint is given by,
\begin{equation}\label{eq:TypeIXFriedConst}
   \bar{\Sigma}^2 +\bar{\Omega} +\bar{V}=1 
\end{equation}
where
\begin{equation*}
    \bar{\Sigma}^2 =  \bar{\Sigma}_+^2 +  \bar{\Sigma}_-^2
\end{equation*}
and
\begin{align*}
    \bar{V} = \frac{1}{12}\left[
    \bar{N}_1^2 +\bar{N}_2 ^2 \bar{N}_3^2 -2\bar{N}_1\bar{N}_2-2\bar{N}_2\bar{N}_3-2\bar{N}_1\bar{N}_3\right.\\
    \left. +3\left(\bar{N}_1\bar{N}_2\bar{N}_3\right)^{2/3}
    \right]
\end{align*}
which implies that $\bar{V}\geq 0$. The corresponding evolution equation for the expansion normalised density parameter is given by,
\begin{equation}
    \bar{\Omega}^{\star} = \left[-(3\gamma -2)\bar{V}+ 3(2-\gamma)\bar{\Sigma}^2\right]-\kappa \bar{H}\bar{\Sigma}^2
\end{equation}
Now using the Friedmann constraint $\bar{q}$ can also be written as,
\begin{equation}\label{eq:defofqbar}
    \bar{q} = 2 \bar{\Sigma}^2 +\frac{1}{2}(3\gamma -2)\bar{\Omega}
\end{equation}
and $\bar{q} \geq 0$ being equal to $0$ when $\bar{\Sigma}^2 =\bar{\Omega}=0$. Using \eqref{eq:defHbar}, we find that $\bar{H}$ is bounded as $-1<\bar{H}<1$. The evolution equation for $\bar{H}$ is given as,
\begin{equation}
    \bar{H}^{\star} = -(1-\bar{H}^2)\bar{q}
\end{equation}
Using the fact that $\bar{q} \geq 0$, and $\vert \bar{H}\vert <1$, we can assume that $ \bar{H}^{\star} <0$. This implies that $\lim _{\bar{\tau}\rightarrow -\infty}\bar{H} = 1$ and $\lim _{\bar{\tau}\rightarrow +\infty}\bar{H} =-1$. This implies that an initially expanding model will undergo a recollapse when $\bar{H}$ will become zero. After the recollapse point i.e.\ $\bar{\tau}\rightarrow \infty$, the condition

\begin{equation}
    \kappa > 3\left(2-\gamma\right)
\end{equation}

along with the positivity of $\bar{\Sigma}^2$ and $\bar{V}$ as well as requiring that the Strong Energy Condition $(\rho +3p)>0$ ensure that $ \bar{\Omega}^{\star} \geq 0$. Thus $\lim _{\bar{\tau}\rightarrow +\infty}\bar{\Omega} =1$ which also implies that $\lim _{\bar{\tau}\rightarrow +\infty}\bar{\Sigma}^2=0$ and $\lim _{\bar{\tau}\rightarrow +\infty}\bar{V}=0$.

From \eqref{eq:defofqbar} then implies that $\lim _{\bar{\tau}\rightarrow +\infty}\bar{q} = (3\gamma -2)/2$. From the evolution equations for the $\bar{N}_i$ for $i=1,2,3$, there exists a parameter $\epsilon >0$ so that, $d\mathrm{ln}\bar{N}_a/d\bar{\tau}<-\epsilon$ and so $\lim _{\bar{\tau}\rightarrow +\infty}\bar{N}_a =0$. Thus we can arrive at a similar conclusion to types I-VIII, that the inclusion of a shear viscous term with the coefficient of viscosity obeying the condition \eqref{eq:condition} will act as an isotropisation mechanism to prevent the growth of anisotropies on approach to the expansion minima or `bounce'.

\section{Bianchi Class B}

For Class B models the parameter A defined in \eqref{eq:paramAB} is non-zero. Following \cite{hwain}, we define the shear variables $\tilde{\sigma}\equiv\frac{1}{2}\tilde{\sigma}_{ab}\tilde{\sigma}^{ab}$ where $\tilde{\sigma}_{ab}$ is the traceless part of the shear tensor $\sigma_{ab}$. The trace for Class A is zero but for Class B is given by $\sigma_+ \equiv \frac{1}{2}\sigma_a^a$. Similarly we can define $\tilde{n}\equiv \frac{1}{6}\tilde{n}_{ab}\tilde{n}^{ab}$ where $\tilde{n}_{ab}$ is the traceless part of $n_{ab}$ and the trace is given by $n_+ =\frac{1}{2}n^a_a$. The quantities $n_+$ and $\tilde{n}$ are related through the non-zero $A$ and the group parameter $h=\tilde{h}^{-1}$ (for some constant $\tilde{h}$) as,
\begin{equation}\label{eq:tildeN}
    \tilde{n} = \frac{1}{3}\left(n_+^2 - \tilde{h}A^2\right)
\end{equation}

We proceed with the expansion normalisation as in the case of Types I-VIII in Class A,
\begin{equation}
 \left(\Sigma_+,\tilde{\Sigma},N_+,\tilde{N},\Omega,\tilde{A}\right)=\left(\frac{\sigma_+}{H},\frac{\tilde{\sigma}}{H},\frac{n_+}{H},\frac{\tilde{n}}{H},\frac{\rho}{3H^2},\frac{A^2}{H^2}\right)   
\end{equation}
Of course following \eqref{eq:tildeN}, we have,
\begin{equation}
     \tilde{N} = \frac{1}{3}\left(N_+^2 - \tilde{h}\tilde{A}\right)
\end{equation}
Following \cite{Lidsey2005,hwain}, we use the fact that the evolution equation for the expansion normalised variable $\Omega$ for Bianchi Class B remains the same as \eqref{eq:EvModOmega} and the definition of $q$ in \eqref{eq:defOfq} also doesn't change. However now the variables $K$ and $\Sigma ^2$ defined previously in \eqref{eq:defK} and \eqref{eq:defsigma} are still positive definite, and are given by,
\begin{align}
    K= \tilde{N}+\tilde{A}\\
    \Sigma ^2 = \tilde{\Sigma} + \Sigma_+^2 
\end{align}
In correspondence with the condition drawn on $\Omega'\leq0$ in Types I-VIII, we have $\lim_{\tau \rightarrow -\infty}K=\lim_{\tau \rightarrow -\infty}\Sigma ^2 =0$ and by \eqref{eq:Friedmann_constraint}, $\lim_{\tau \rightarrow -\infty}\Omega =1$.
 This further implies that $\lim_{\tau \rightarrow -\infty}\Sigma =0$ and  $\lim_{\tau \rightarrow -\infty}K =0$, that  $\lim_{\tau \rightarrow -\infty}\tilde{\Sigma} =0$ and  $\lim_{\tau \rightarrow -\infty}\Sigma_+ =0$. And that $\lim_{\tau \rightarrow -\infty}\tilde{N} =0$ and  $\lim_{\tau \rightarrow -\infty}\tilde{A} =0$. Thus the spatial curvature variables given by,
 \begin{equation}
     \mathcal{S}_+ = 2\tilde{N},\;\;\;\;\;\;\mathcal{S}^{ab}\mathcal{S}_{ab} = 24\left(\tilde{A}+\tilde{N}^2\right)\tilde{N}
 \end{equation}
 
 vanish at the bounce. The conclusions drawn about the stability of the isotropic FL point in an initially contracting Universe with an equation of state $p=\left(\gamma(\rho)-1\right)\rho$ such that $\rho +3p>0$ drawn in Bianchi Types I-VIII remain unchanged.
\newline
\section{Conclusion}

The problem of isotropisation in contracting Universes has been a key challenge in the formulation of bouncing alternatives to inflation. In \cite{me4}, we have shown that the inclusion of shear viscous stresses leads to isotropisation of a Bianchi Type IX Universe. In this work we have used the orthonormal frame formalism used by \cite{Belinskii1972,WEllis} and subsequent expansion normalisation to write the Einstein Field Equations as a phase plane system. This has allowed us to derive global stability theorems for the Bianchi Class I-VIII and separately Type IX as well as the Bianchi Class B, which show that the isotropic, spatially homoegenous solution becomes the attractor in an initially contracting Universe on approach to the expansion minima. 
\paragraph*{}
Our results hold for an equation of state following a form given by $p=\left(\gamma(\rho)-1\right)\rho$ and hence the case of non-linear equation of state is encapsulated in our proof. In the case of positive spatial curvature, with a Universe sourced with the non-linear equation of state fluid \cite{bruni1,bruni2,me4}, the expansion minima results in a `bounce' and a subsequent re-expansion. In this work we have demonstrated only the isotropisation in the contracting phase of the most general spatially homogeneous cosmology and have not modelled the physics of a bounce.
\paragraph*{}
The inclusion of shear viscosity in the contracting phase thus appears to be an effective isotropisation mechanism for all types of spatially homogeneous cosmologies - without having to use an ekpyrotic fluid $p\gg\rho$ as is done in \cite{Lidsey2005} . As the form of shear viscosity used here and in \cite{me4} is dependent on the Hubble parameter $H$ through $\pi_{\alpha\beta}=-\kappa H\sigma_{\alpha\beta}$, this viscous effect will vanish at the bounce point $H=0$. This should have an effect on the damping of perturbative anisotropies in the contracting phase but should not produce significant gravitational waves at the bounce. 
\paragraph*{}
The effect of fully non-linear inhomogeneities is yet to be studied in these models. There have been some numerical studies of its evolution in the contracting phase with an ekpyrotic phase \cite{will1,will2}. At the linear level, the scalar perturbations are dependent on the shear tensor through the electric part of the Weyl Tensor and hence should remain small in the case of effective isotropisation. Furthermore, if we follow the assumptions made in \cite{bkl,belinski_henneaux_2017}, then on approach to a singularity, each point of an inhomogeneous Universe behaves as a Bianchi cosmology and hence these results should hold. However full numerical GR should be used to test whether this is actually the case.

\appendix*

\section{Vanishing of the Fermi rotation}

The Fermi rotation, which is defined as the local angular velocity of the spatial frame specified by a basis of one-forms with respect to the Fermi propagated frame is given by,

\begin{equation}
    \Omega ^{\alpha} = \frac{1}{2}\epsilon^{\alpha \mu \nu}e^i_{\mu}e_{\nu i;j}u^j
\end{equation}

We have seen that it is a term in the shear evolution equations\eqref{eq:evshear}. This term vanishes for vanishining anisotropic stress or for anisotropic stress that takes the form of shear viscosity i.e.\ $\pi_{\alpha \beta} \propto \sigma_{\alpha \beta}$.
To demonstrate this, let us consider a term of the form,
\begin{equation}
    \epsilon ^{\mu \nu}_{(\alpha}\sigma _{\beta)\mu}
\end{equation}
This term shows up in the evolution equation for the shear tensor $\sigma_{\alpha \beta}$,
Now as $\alpha = \beta$ as the LHS of the equation is $\dot{\sigma} _{\alpha \beta}$ and we have diagonalised $\sigma _{\alpha \beta}$. The RHS has terms where at least $2$ indices are equal of the Levi Civita symbol $\epsilon^{\mu \nu}_{\alpha}$ when you expand out the sum. So both the terms of the form $\epsilon^{\mu \nu}_{(\alpha}\sigma_{\beta)\nu}$ and $\epsilon^{\mu \nu}_{(\alpha}n_{\beta)\nu}$ are zero if $\sigma _{\alpha \beta}$ and $n_{\alpha \beta}$ are diagonalised. So we have the freedom to set $\Omega _{\nu} =0$.
\paragraph*{}
So now let's check if we can diagonalise $\sigma_{\alpha \beta}$ given that $n_{\alpha \beta}$ is diagonal and $A_{\alpha} =0$ for the case of Class A models. Also we are assuming no momentum transfer in our model so $q_{\alpha} =0$. Now $q_{\alpha}$ is given by,
\begin{equation}
    q_{\alpha} = 3 \sigma _{\alpha}^{\mu}a_{\mu} -\epsilon ^{\mu \nu}_{\alpha}\sigma_{\mu}^{\beta}n_{\beta \nu}
\end{equation}
So $\epsilon ^{\mu \nu}_{\alpha}\sigma_{\mu}^{\beta}n_{\beta \nu} = 0$ and now $\beta = \nu$ for all $n_{\beta \nu} \neq 0$. By expanding out the sum index by index, we can see that for this term to be zero, all the terms where $\sigma _{\alpha \beta}$ with $\alpha \neq \beta$ have to be zero. Let us look at the case of $\alpha = \beta = 1$ as an example.
\begin{equation}
    \epsilon ^{\mu 1}_{\alpha}\sigma ^1_{\mu}n_{11} = \epsilon ^{01}_{\alpha}\sigma ^1_0 n_{11} + \epsilon ^{11}_{\alpha}\sigma ^1_1 n_{11}  + \epsilon ^{21}_{\alpha}\sigma ^1_2 n_{11}  + \epsilon ^{31}_{\alpha}\sigma ^1_3 n_{11} 
\end{equation}
 The term that has $2$ indices equal in the Levi Civita term is zero. All the other terms have to be zero to satisfy the no momentum transfer ($q_{\alpha}=0$) and the $a_{\alpha} =0$ conditions. From the above, if the cross terms of $\sigma _{\alpha \beta}$ cannot vanish, then they can be expressed in terms of one another. A solution to this system of equations is for them to be individually zero. So we can diagnolise $\sigma _{\alpha \beta}$ if $n_{\alpha \beta}$ is diagonal. The other aspect of this is to ensure that if we start out with a diagonal $\sigma _{\alpha \beta}$, there are no off diagonal terms in $\dot{\sigma}_{\alpha \beta}$ that mean that after an initial time some cross terms of $\sigma _{\alpha \beta}$ become non zero. This is most easily achieved in the case of no anisotropic stress. In the case of general anisotropic stress, it is no longer true that choosing a frame where $n_{\alpha \beta}$ is diagonal implies that  $\sigma _{\alpha \beta}$ is also diagonal. But in our case, with shear viscosity, $\pi _{\alpha \beta}$ is proportional to  $\sigma _{\alpha \beta}$ and so we can still diagonalise it. Hence in our case, we have the freedom to set the Fermi rotation $\Omega _{\alpha} =0$.

 \section{Limits on the shear viscous entropy}
 
 The shear viscosity can be used to model anisotropy dissipative processes up to a certain limit. There is an upper limit on the amount of entropy production from shear viscous stresses. This limit can be calculated for example at the era of grand unification if collisionless processes are ignored. We present the calculation that has been worked out in detail in \cite{barrow_dis_uni}.
 We define an anisotropy factor as, \cite{barrow_dis_uni},
 \begin{equation}
     A = \left(\frac{\rho_{\beta}}{\rho}\right)^{1/2}
 \end{equation}
 where the energy density in the anisotropies is given by,
 \begin{equation}
     \rho_{\beta} \sim \sigma_{ab} \sigma^{ab}
 \end{equation}
 The maximum anisotropic state of a cosmology is given by the Kasner geometry where the expansion rate is given by,
 \begin{equation}
     H = \frac{1}{3t}
 \end{equation}
 Using the Friedmann equation at anisotropy domination,
 \begin{equation}
     3 M_{Pl}^2 H^2 =\rho _{\beta}
 \end{equation}
 we can find the maximum value of the anisotropy parameter to be,
 \begin{equation}
     A_{max}= \frac{g_{\star}^{1/2}\alpha_S^2 M_{Pl}}{4\pi^{3/2}\sqrt{80}T_i}
 \end{equation}
 where $T_i$ is the temperature at which dissipation occurs when the Universe cools.
 The entropy per baryon $\Delta$ that can be extracted is then given by,
 \begin{equation}
     \Delta \sim \left(1+\frac{A}{4A_{max}}\right)^3
 \end{equation}
 In order to find the maximum upper bound we replace $A$ by $A_{max}$, and find this to be $\frac{125}{64}$.
 If the anisotropy were to exceed $A_{max}$, then the anisotropy damping would be collisionless and we would not be able to use the viscosity description anymore. The viscosity description is only applicable when $t_{coll} < t_{exp}$, where $t_{coll}$ and $t_{exp}$ are the collision times of the particle interactions and the expansion time of the Universe respectively. The assumption in the above result is that the interaction cross section is given by that of the Grand Unified Interaction $\sigma \sim \alpha_S^2 T^{-2}$.

 \acknowledgments

CG would like to thank John D. Barrow for comments on the manuscript, and Marco Bruni for discussions on an earlier version of the work. She would also like to thank Wolfson College and the Cambridge Philosophical Society for the Henslow Fellowship which has supported this work.

\end{document}